\documentclass[%
 reprint,
superscriptaddress,
nofootinbib,
 amsmath,amssymb,prd,showpacs
 aps,
]{revtex4-2}

\usepackage{graphicx}
\usepackage{dcolumn}
\usepackage{bm}
\usepackage{siunitx}
\usepackage{xspace}
\usepackage{booktabs}
\usepackage{hyperref}
\usepackage[capitalize]{cleveref}
\usepackage{xcolor}
\makeatletter
\renewcommand*{\@fnsymbol}[1]{\ensuremath{\ifcase#1\or *\or \dagger\or \ddagger\or
   \mathsection\or \mathparagraph\or \|\or **\or \dagger\dagger
   \or \ddagger\ddagger \or \mathparagraph\mathparagraph \else\@ctrerr\fi}}
\makeatother

\usepackage[normalem]{ulem}
\providecommand{\ssqthonethree}{\ensuremath{\sin^2\theta_{13}}\xspace}
\providecommand{\ssqthtwothree}{\ensuremath{\sin^2\theta_{23}}\xspace}
\providecommand{\ssqthonetwo}{\ensuremath{\sin^2\theta_{12}}\xspace}

\providecommand{\deltacp}{\ensuremath{\delta_{\scriptscriptstyle\mathrm{CP}}}\xspace}

\providecommand{\dmsqtwothree}{\ensuremath{\Delta{}m^2_{32}}\xspace}
\providecommand{\dmsqtwothreeb}{\ensuremath{\Delta{}\bar{m}^2_{32}}\xspace}
\providecommand{\dmsqonetwo}{\ensuremath{\Delta{}m^2_{21}}\xspace}

\providecommand{\nubar}{\ensuremath{\bar{\nu}}\xspace}
\providecommand{\nue}{\ensuremath{\nu_{e}}\xspace}
\providecommand{\numu}{\ensuremath{\nu_{\mu}}\xspace}

\providecommand{\numub}{\ensuremath{\nubar_{\mu}}\xspace}

\def\thatm   {\ensuremath{\theta_{23}}\xspace}
\def\thatmb  {\ensuremath{\bar{\theta}_{23}}\xspace}

\newlength{\parenbarKernelHeight}
\providecommand{\parenbar}[1]{%
    \settoheight{\parenbarKernelHeight}{\ensuremath{#1}}%
    \addtolength{\parenbarKernelHeight}{0.3pt}
    \llap{\raisebox{\parenbarKernelHeight}{\scalebox{0.5}[0.5]{(}}}%
    \overline{#1}%
    \rlap{\raisebox{\parenbarKernelHeight}{\scalebox{0.5}[0.5]{)}}}%
}

\providecommand{\CCzeropi}{CC-$0\pi$\xspace}
\providecommand{\CConepiplus}{CC-$1\pi^+$\xspace}

\providecommand{\CCother}{CC-Other\xspace}

\begin{document}

\title{Updated T2K measurements of muon neutrino and antineutrino disappearance using 3.6 $\times$ 10$^{21}$ protons on target}


\newcommand{\INSTHD}{\affiliation{University Autonoma Madrid, Department of Theoretical Physics, 28049 Madrid, Spain}}
\newcommand{\INSTEE}{\affiliation{University of Bern, Albert Einstein Center for Fundamental Physics, Laboratory for High Energy Physics (LHEP), Bern, Switzerland}}
\newcommand{\INSTFE}{\affiliation{Boston University, Department of Physics, Boston, Massachusetts, U.S.A.}}
\newcommand{\INSTGA}{\affiliation{University of California, Irvine, Department of Physics and Astronomy, Irvine, California, U.S.A.}}
\newcommand{\INSTI}{\affiliation{IRFU, CEA, Universit\'e Paris-Saclay, F-91191 Gif-sur-Yvette, France}}
\newcommand{\INSTGB}{\affiliation{University of Colorado at Boulder, Department of Physics, Boulder, Colorado, U.S.A.}}
\newcommand{\INSTFG}{\affiliation{Colorado State University, Department of Physics, Fort Collins, Colorado, U.S.A.}}
\newcommand{\INSTFH}{\affiliation{Duke University, Department of Physics, Durham, North Carolina, U.S.A.}}
\newcommand{\INSTJA}{\affiliation{E\"{o}tv\"{o}s Lor\'{a}nd University, Department of Atomic Physics, Budapest, Hungary}}
\newcommand{\INSTEF}{\affiliation{ETH Zurich, Institute for Particle Physics and Astrophysics, Zurich, Switzerland}}
\newcommand{\INSTIE}{\affiliation{CERN European Organization for Nuclear Research, CH-1211 Gen\'eve 23, Switzerland}}
\newcommand{\INSTEG}{\affiliation{University of Geneva, Section de Physique, DPNC, Geneva, Switzerland}}
\newcommand{\INSTHJ}{\affiliation{University of Glasgow, School of Physics and Astronomy, Glasgow, United Kingdom}}
\newcommand{\INSTDG}{\affiliation{H. Niewodniczanski Institute of Nuclear Physics PAN, Cracow, Poland}}
\newcommand{\INSTCB}{\affiliation{High Energy Accelerator Research Organization (KEK), Tsukuba, Ibaraki, Japan}}
\newcommand{\INSTIB}{\affiliation{University of Houston, Department of Physics, Houston, Texas, U.S.A.}}
\newcommand{\INSTED}{\affiliation{Institut de Fisica d'Altes Energies (IFAE) - The Barcelona Institute of Science and Technology, Campus UAB, Bellaterra (Barcelona) Spain}}
\newcommand{\INSTJC}{\affiliation{Institut f\"ur Physik, Johannes Gutenberg-Universit\"at Mainz, Staudingerweg 7, 55128 Mainz, Germany}}
\newcommand{\INSTEC}{\affiliation{IFIC (CSIC \& University of Valencia), Valencia, Spain}}
\newcommand{\INSTHH}{\affiliation{Institute For Interdisciplinary Research in Science and Education (IFIRSE), ICISE, Quy Nhon, Vietnam}}
\newcommand{\INSTEI}{\affiliation{Imperial College London, Department of Physics, London, United Kingdom}}
\newcommand{\INSTGF}{\affiliation{INFN Sezione di Bari and Universit\`a e Politecnico di Bari, Dipartimento Interuniversitario di Fisica, Bari, Italy}}
\newcommand{\INSTBE}{\affiliation{INFN Sezione di Napoli and Universit\`a di Napoli, Dipartimento di Fisica, Napoli, Italy}}
\newcommand{\INSTBF}{\affiliation{INFN Sezione di Padova and Universit\`a di Padova, Dipartimento di Fisica, Padova, Italy}}
\newcommand{\INSTBD}{\affiliation{INFN Sezione di Roma and Universit\`a di Roma ``La Sapienza'', Roma, Italy}}
\newcommand{\INSTEB}{\affiliation{Institute for Nuclear Research of the Russian Academy of Sciences, Moscow, Russia}}
\newcommand{\INSTHI}{\affiliation{International Centre of Physics, Institute of Physics (IOP), Vietnam Academy of Science and Technology (VAST), 10 Dao Tan, Ba Dinh, Hanoi, Vietnam}}
\newcommand{\INSTJD}{\affiliation{ILANCE, CNRS – University of Tokyo International Research Laboratory, Kashiwa, Chiba 277-8582, Japan}}
\newcommand{\INSTHA}{\affiliation{Kavli Institute for the Physics and Mathematics of the Universe (WPI), The University of Tokyo Institutes for Advanced Study, University of Tokyo, Kashiwa, Chiba, Japan}}
\newcommand{\INSTID}{\affiliation{Keio University, Department of Physics, Kanagawa, Japan}}
\newcommand{\INSTIF}{\affiliation{King's College London, Department of Physics, Strand, London WC2R 2LS, United Kingdom}}
\newcommand{\INSTCC}{\affiliation{Kobe University, Kobe, Japan}}
\newcommand{\INSTCD}{\affiliation{Kyoto University, Department of Physics, Kyoto, Japan}}
\newcommand{\INSTEJ}{\affiliation{Lancaster University, Physics Department, Lancaster, United Kingdom}}
\newcommand{\INSTII}{\affiliation{Lawrence Berkeley National Laboratory, Berkeley, CA 94720, USA}}
\newcommand{\INSTBA}{\affiliation{Ecole Polytechnique, IN2P3-CNRS, Laboratoire Leprince-Ringuet, Palaiseau, France}}
\newcommand{\INSTFC}{\affiliation{University of Liverpool, Department of Physics, Liverpool, United Kingdom}}
\newcommand{\INSTFI}{\affiliation{Louisiana State University, Department of Physics and Astronomy, Baton Rouge, Louisiana, U.S.A.}}
\newcommand{\INSTIH}{\affiliation{Joint Institute for Nuclear Research, Dubna, Moscow Region, Russia}}
\newcommand{\INSTHB}{\affiliation{Michigan State University, Department of Physics and Astronomy,  East Lansing, Michigan, U.S.A.}}
\newcommand{\INSTCE}{\affiliation{Miyagi University of Education, Department of Physics, Sendai, Japan}}
\newcommand{\INSTDF}{\affiliation{National Centre for Nuclear Research, Warsaw, Poland}}
\newcommand{\INSTFJ}{\affiliation{State University of New York at Stony Brook, Department of Physics and Astronomy, Stony Brook, New York, U.S.A.}}
\newcommand{\INSTGJ}{\affiliation{Okayama University, Department of Physics, Okayama, Japan}}
\newcommand{\INSTCF}{\affiliation{Osaka Metropolitan University, Department of Physics, Osaka, Japan}}
\newcommand{\INSTGG}{\affiliation{Oxford University, Department of Physics, Oxford, United Kingdom}}
\newcommand{\INSTIC}{\affiliation{University of Pennsylvania, Department of Physics and Astronomy,  Philadelphia, PA, 19104, USA.}}
\newcommand{\INSTGC}{\affiliation{University of Pittsburgh, Department of Physics and Astronomy, Pittsburgh, Pennsylvania, U.S.A.}}
\newcommand{\INSTFA}{\affiliation{Queen Mary University of London, School of Physics and Astronomy, London, United Kingdom}}
\newcommand{\INSTE}{\affiliation{University of Regina, Department of Physics, Regina, Saskatchewan, Canada}}
\newcommand{\INSTGD}{\affiliation{University of Rochester, Department of Physics and Astronomy, Rochester, New York, U.S.A.}}
\newcommand{\INSTHC}{\affiliation{Royal Holloway University of London, Department of Physics, Egham, Surrey, United Kingdom}}
\newcommand{\INSTBC}{\affiliation{RWTH Aachen University, III. Physikalisches Institut, Aachen, Germany}}
\newcommand{\INSTJB}{\affiliation{Departamento de F\'isica At\'omica, Molecular y Nuclear, Universidad de Sevilla, 41080 Sevilla, Spain}}
\newcommand{\INSTFB}{\affiliation{University of Sheffield, Department of Physics and Astronomy, Sheffield, United Kingdom}}
\newcommand{\INSTDI}{\affiliation{University of Silesia, Institute of Physics, Katowice, Poland}}
\newcommand{\INSTBB}{\affiliation{Sorbonne Universit\'e, Universit\'e Paris Diderot, CNRS/IN2P3, Laboratoire de Physique Nucl\'eaire et de Hautes Energies (LPNHE), Paris, France}}
\newcommand{\INSTEH}{\affiliation{STFC, Rutherford Appleton Laboratory, Harwell Oxford,  and  Daresbury Laboratory, Warrington, United Kingdom}}
\newcommand{\INSTCH}{\affiliation{University of Tokyo, Department of Physics, Tokyo, Japan}}
\newcommand{\INSTBJ}{\affiliation{University of Tokyo, Institute for Cosmic Ray Research, Kamioka Observatory, Kamioka, Japan}}
\newcommand{\INSTCG}{\affiliation{University of Tokyo, Institute for Cosmic Ray Research, Research Center for Cosmic Neutrinos, Kashiwa, Japan}}
\newcommand{\INSTHF}{\affiliation{Tokyo Institute of Technology, Department of Physics, Tokyo, Japan}}
\newcommand{\INSTGI}{\affiliation{Tokyo Metropolitan University, Department of Physics, Tokyo, Japan}}
\newcommand{\INSTHG}{\affiliation{Tokyo University of Science, Faculty of Science and Technology, Department of Physics, Noda, Chiba, Japan}}
\newcommand{\INSTF}{\affiliation{University of Toronto, Department of Physics, Toronto, Ontario, Canada}}
\newcommand{\INSTB}{\affiliation{TRIUMF, Vancouver, British Columbia, Canada}}
\newcommand{\INSTDJ}{\affiliation{University of Warsaw, Faculty of Physics, Warsaw, Poland}}
\newcommand{\INSTDH}{\affiliation{Warsaw University of Technology, Institute of Radioelectronics and Multimedia Technology, Warsaw, Poland}}
\newcommand{\INSTIJ}{\affiliation{Tohoku University, Faculty of Science, Department of Physics, Miyagi, Japan}}
\newcommand{\INSTFD}{\affiliation{University of Warwick, Department of Physics, Coventry, United Kingdom}}
\newcommand{\INSTGH}{\affiliation{University of Winnipeg, Department of Physics, Winnipeg, Manitoba, Canada}}
\newcommand{\INSTEA}{\affiliation{Wroclaw University, Faculty of Physics and Astronomy, Wroclaw, Poland}}
\newcommand{\INSTHE}{\affiliation{Yokohama National University, Department of Physics, Yokohama, Japan}}
\newcommand{\INSTH}{\affiliation{York University, Department of Physics and Astronomy, Toronto, Ontario, Canada}}

\INSTHD
\INSTEE
\INSTFE
\INSTGA
\INSTI
\INSTGB
\INSTFG
\INSTFH
\INSTJA
\INSTEF
\INSTIE
\INSTEG
\INSTHJ
\INSTDG
\INSTCB
\INSTIB
\INSTED
\INSTJC
\INSTEC
\INSTHH
\INSTEI
\INSTGF
\INSTBE
\INSTBF
\INSTBD
\INSTEB
\INSTHI
\INSTJD
\INSTHA
\INSTID
\INSTIF
\INSTCC
\INSTCD
\INSTEJ
\INSTII
\INSTBA
\INSTFC
\INSTFI
\INSTIH
\INSTHB
\INSTCE
\INSTDF
\INSTFJ
\INSTGJ
\INSTCF
\INSTGG
\INSTIC
\INSTGC
\INSTFA
\INSTE
\INSTGD
\INSTHC
\INSTBC
\INSTJB
\INSTFB
\INSTDI
\INSTBB
\INSTEH
\INSTCH
\INSTBJ
\INSTCG
\INSTHF
\INSTGI
\INSTHG
\INSTF
\INSTB
\INSTDJ
\INSTDH
\INSTIJ
\INSTFD
\INSTGH
\INSTEA
\INSTHE
\INSTH

\author{K.\,Abe}\INSTBJ
\author{N.\,Akhlaq}\INSTFA
\author{R.\,Akutsu}\INSTCB
\author{A.\,Ali}\INSTGH\INSTB
\author{S.\,Alonso Monsalve}\INSTEF
\author{C.\,Alt}\INSTEF
\author{C.\,Andreopoulos}\INSTFC
\author{M.\,Antonova}\INSTEC
\author{S.\,Aoki}\INSTCC
\author{T.\,Arihara}\INSTGI
% ...

\collaboration{The T2K Collaboration}\noaffiliation

\date{\today}

\begin{abstract}

Muon neutrino and antineutrino disappearance probabilities are identical in the standard three-flavor neutrino oscillation framework, but \textit{CPT} violation and nonstandard interactions can violate this symmetry. In this work we report the measurements of $\sin^{2} \theta_{23}$ and $\Delta m_{32}^2$ independently for neutrinos and antineutrinos. The aforementioned symmetry violation would manifest as an inconsistency in the neutrino and antineutrino oscillation parameters. The analysis discussed here uses a total of 1.97$\times$10$^{21}$ and 1.63$\times$10$^{21}$ protons on target taken with a neutrino and antineutrino beam respectively, and benefits from improved flux and cross section models, new near-detector samples and more than double the data reducing the overall uncertainty of the result. No significant deviation is observed, consistent with the standard neutrino oscillation picture.

\end{abstract}

\maketitle

\section{Introduction}
\label{sec:intro}

Neutrino oscillations are described by the Pontecorvo-Maki-Nakagawa-Sakata (PMNS) matrix, $U_{lj}$, which relates the neutrino mass eigenstates $\nu_j$ [with masses $m_j=(m_1,m_2,m_3)$] to the left-handed neutrino flavor fields $\nu_{l}$  ($\nu_e$, $\nu_{\mu}$, $\nu_{\tau}$)\cite{Pontecorvo:1967fh,MNS10.1143/PTP.28.870} as $ \nu_l = \sum_j U_{lj} \nu_j$. The matrix $U_{lj}$ is parametrized by three mixing angles $\theta_{12}$, $\theta_{13}$, and $\theta_{23}$, and a \textit{CP}-violating phase \deltacp. Two Majorana phases appear on the diagonal terms in $U_{lj}$ if the neutrino is the same as its antiparticle, but they have no effect on neutrino oscillations. In this framework, \numu and \numub disappearance probabilities are the same in the absence of matter effects (which are negligible at T2K energies and baseline, but are included in their calculation) so a mismatch could indicate a source of \textit{CPT} violation (since $CPT[P(\numu \rightarrow \numu)] = P(\numub \rightarrow \numub)$ in vacuum) or a source of nonstandard interactions\cite{Arguelles:2022xxa}. 

The results presented in this paper represent an update to the previous T2K measurements~\cite{PhysRevD.103.L011101,PhysRevD.96.011102,PhysRevLett.116.181801}. Like these previous analyses, we allow the oscillation parameters for \numu (\thatm, \dmsqtwothree) to vary separately from those of \numub (\thatmb, \dmsqtwothreeb), while all other oscillation parameters are assumed to be the same for neutrinos and antineutrinos.

This work is organized as follows. First, an overview of the T2K experimental setup is given in \cref{sec:t2k}. The analysis method is then described in \cref{sec:analysis}. Finally, the results are discussed in \cref{sec:results} and conclusions are presented in \cref{sec:conclusions}.

\section{T2K experimental setup}
\label{sec:t2k}

T2K is a long-baseline neutrino oscillation experiment located in Japan~\cite{ABE2011106}. A neutrino beam produced at the Japan Proton Accelerator Research Complex (J-PARC) is directed towards Super-Kamiokande (SK)~\cite{FUKUDA2003418,ABE2014253}, a large water Cherenkov detector. 

The neutrino beam is produced by \SI{30}{\GeV} protons impinging on a graphite target. Interactions in the target produce hadrons, which are focused using three magnetic horns~\cite{SEKIGUCHI201557}. The polarity of the magnetic field produced by the horns is reversible, allowing for the selection of positively (negatively) charged hadrons which then decay into a beam dominated by muon neutrinos (antineutrinos).

A suite of near detectors is situated \SI{280}{\m} downstream of the beam production target. The stability and direction of the neutrino beam are monitored using the on-axis near detector INGRID~\cite{Abe:2011xv}. INGRID consists of 14 detector modules arranged in a cross formation, with each module containing sandwiched layers of iron plates and scintillator planes~\cite{OTANI2010368}. A second near detector, ND280, is positioned \SI{2.5}{\degree} off-axis from the neutrino beamline. It is used to measure the unoscillated neutrino flux and neutrino interaction parameters in order to constrain systematic errors in the oscillation analysis. ND280 consists of a $\pi^{0}$ detector~\cite{ASSYLBEKOV201248} followed by three time-projection chambers (TPCs)~\cite{ABGRALL201125} interleaved with two fine-grained detectors (FGDs)~\cite{T2KND280FGD:2012umz}, all surrounded by an electromagnetic calorimeter~\cite{Allan_2013}. ND280 is also magnetized to allow for the charge identification of particles. The gaps in the magnet yoke are instrumented by muon range detectors~\cite{AOKI2013135}. 

SK is a 50 kt water Cherenkov detector situated \SI{295}{\km} downstream of the neutrino production point and is positioned at the same off-axis angle as ND280. In this configuration, the beam has a peak energy around \SI{0.6}{\GeV} that maximizes the effect of neutrino oscillations. It has optically separated inner detector (ID) and outer detector (OD) volumes. It uses 11,129 inward-facing 20-inch photomultiplier tubes (PMTs) to detect Cherenkov radiation from charged particles traversing the detector. To reject interactions from outside the ID volume, 1,885 outward-facing 8-inch PMTs in the OD are used. SK is able to discriminate between electrons and muons by their Cherenkov ring profiles~\cite{Super-Kamiokande:2019gzr}.

\section{Analysis method}
\label{sec:analysis}

The analysis strategy presented here is similar to the one employed in previous analyses~\cite{PhysRevD.103.L011101,PhysRevD.96.011102,PhysRevLett.116.181801}. First, we define a model that predicts the event spectra at both the near and far detectors. Such predictions are extracted by simulating the neutrino flux and cross sections, tuned to external experimental data, and the detector response. This model is then fit to the ND280 data to obtain tuned values and constraints for the flux systematic uncertainties and a subset of the cross section systematic uncertainties. The results of the near-detector analysis are propagated to SK as a multivariate normal distribution described by a covariance matrix and the best-fit values for each parameter associated to neutrino flux and cross section systematic uncertainties. At this point, we perform a fit to SK data to extract the oscillation parameters. Four significant updates have been made since the previous analysis. First, the number of protons on target (POT) collected in neutrino beam mode was increased from $1.49 \times 10^{21}$ to $1.97 \times 10^{21}$ by including T2K data up to February 2020. Second, the flux prediction was tuned to the $\pi^\pm$ yields from the surface of a T2K replica target measured by NA61/SHINE~\cite{Abgrall:2016jif}. Third, the modeling of neutrino interactions on nuclear targets was improved. Finally, the selection of antineutrino events at ND280 was refined and the data set doubled.

\subsection{Flux prediction}
\label{sec:flux}

The neutrino flux prediction used for this analysis has been upgraded from a tuning~\cite{T2K:2021xwb,Abe:2012av} based on thin target measurements~\cite{Abgrall:2015hmv} to a tuning of charged pion yields~\cite{Abgrall:2016jif} measured by NA61/SHINE using a replica of the T2K target. The details of the new tuning are described in Ref.~\cite{T2K:2023smv}, which is also summarized below.

Incoming protons are generated according to beam profiles measured for each run, and their hadronic interactions inside the $90~\mathrm{cm}$ long graphite target are simulated wìth FLUKA version 2011.2x~\cite{Ahdida:2022gjl,BATTISTONI201510}. The particles emitted from the target are then focused by the three magnetic horns and tracked until they decay into neutrinos in the decay volume using the GEANT3-based \textsc{Jnubeam} package~\cite{Abe:2012av}. Charged pions exiting from the target are tuned using tuning factors based on replica target measurements, which depend on the exiting longitudinal position, momentum, and direction with respect to the target axis. For exiting particles not covered by the replica target measurements, such as kaons and protons and any hadronic interactions outside of the target, cross section, and multiplicity tuning based on thin target measurements is applied to each interaction as in previous analyses. The statistical and systematic uncertainties on the NA61/SHINE measured yields are then propagated to the flux to estimate the uncertainty on the hadron interactions. For interactions unconstrained by external data, uncertainties are assigned based on comparisons between Monte Carlo (MC) hadron interaction models. Together with other uncertainties on proton beam profile parameters and beamline alignment, a covariance matrix of the flux at the near and far detectors for each neutrino flavor in the two beam modes is constructed. This is then used to propagate the neutrino flux constraint at the near detector to the far detector prediction.

The new tuning, extrapolated using the NA61/SHINE 2009 replica target data, reduces the relative uncertainty of the \numu flux in $\nu$-mode and \numub flux in \nubar-mode from about 9--12\% to 5--8\% near the flux peak. For the \numub component in $\nu$-mode and \numu component in \nubar-mode (the so-called ``wrong-sign background"), the uncertainty has a larger contribution from interactions occurring outside the main target, resulting in a relative uncertainty of about 6--8\%.

\subsection{Neutrino interaction modeling}
\label{sec:nuint}

Neutrino and antineutrino interactions are simulated using the MC event generator NEUT version 5.4.0~\cite{Hayato:2021heg}. The main interaction channels in the range of energies relevant for T2K are: charged-current quasielastic scattering (CCQE), 2p2h (``two particle, two hole") interactions, resonant pion production (RES), and deep inelastic scattering (DIS). 2p2h interactions occur when neutrinos interact with correlated pairs of nucleons, ejecting both from the nucleus. Furthermore, hadrons produced in neutrino interactions on nuclei can interact with the nuclear medium, undergoing so-called final state interactions (FSI). CCQE interactions are simulated according to the Llewellyn-Smith formalism~\cite{LlewellynSmith:1971uhs} with a dipole axial form factor and BBBA05 vector form factors~\cite{Bradford:2006yz}. In this analysis, we moved from the relativistic Fermi gas (RFG) nuclear model to the spectral function (SF) model described in Ref.~\cite{Benhar:1994hw}, with an axial mass $M_A^\mathrm{QE} = 1.03 \:\mathrm{GeV}$ tuned to bubble chamber data~\cite{Stowell:2016jfr,Bernard:2001rs}. The 2p2h interactions are simulated according to the Valencia model described in Ref.~\cite{Nieves:2011pp}. The model for RES is based on the Rein--Sehgal model~\cite{Berger:2007rq} for events with an invariant hadronic mass $W \leq 2 \, \mathrm{GeV}$ (natural units are used throughout the paper), with updated nucleon form factors~\cite{Graczyk:2007bc}. The DIS interaction is calculated for events with invariant hadronic mass  W $>$ 1.3 GeV, using GRV98 parton distribution functions~\cite{Gluck:1998xa} with Bodek-Yang corrections~\cite{Bodek:2003wc}. For 1.3 GeV $<$ W $<$ 2 GeV, only DIS interactions that produce more than one pion are simulated to avoid double counting with the nonresonant single pion production. For values of $W \leq 2 \, \mathrm{GeV}$ a custom hadronization \cite{Bronner:2016huz} is employed, whilst for $W > 2 \, \mathrm{GeV}$ \textsc{Pythia/JetSet}~\cite{Sjostrand:1993yb} is used. Pion FSIs are simulated using a semiclassical intranuclear cascade model by Salcedo and Oset~\cite{Bertini:1972vz,Oset:1986sy}, tuned to recent $\pi^{\pm}$-nucleus scattering data~\cite{PinzonGuerra:2018rju}. Nucleon FSIs are described in an analogous cascade model~\cite{Hayato:2021heg}. The Coulomb interaction between the outgoing charged lepton and the nucleus is implemented as a nucleus- and lepton- flavor-dependent shift in the momentum of the outgoing lepton. The size of such a shift has been determined from an analysis of electron scattering data to be $\sim\pm5\:\mathrm{MeV/}c$~\cite{Gueye:1999mm}. Every parameter relevant to the particular channel described above has uncertainties associated to it. The parametrization employed and such uncertainties are often driven by theory, but additional empirically driven parameters are used since the first alone cannot describe the available neutrino cross section data. Important changes compared to the previous analysis are a new treatment of the removal energy for CCQE interactions, the freedom to change the CCQE cross-section normalization as a function of the momentum transferred, and improved FSIs uncertainties.

Contrary to the Fermi-gas models, the SF model does not have a fixed value for the nuclear binding energy and it can be varied as a parameter. The removal energy shifts are encoded in four parameters depending on whether they affect initial-state protons or neutrons, and if the target is carbon or oxygen. These parameters shift the outgoing lepton momentum of a CCQE interaction and depend on the lepton kinematics, neutrino energy, and flavor.

Recent measurements of the charged-current interactions without mesons in the final state performed by MINER$\nu$A~\cite{Ruterbories:2018gub,Rodrigues:2015hik} and T2K~\cite{Abe:2020jbf,Abe:2020uub} show a clear suppression at low-$Q^2$. In previous T2K analyses that used the Fermi-gas model~\cite{T2K:2023smv} this suppression is achieved by including a nuclear screening effect using the random phase approximation (RPA)~\cite{Nieves:2011yp}. Since the SF model employed in this analysis does not include this suppression, five unconstrained parameters that alter the normalization of the CCQE cross section in the range $Q^2=\{0,0.25\}~\text{GeV}^2$ were included. This range is split into subranges of $0.05~\text{GeV}^2$. For values of the momentum transferred larger than 0.25 GeV$^2$ three parameters are used to account for deviation from the dipole model. 

Finally, the NEUT pion cascade model has been tuned to external $\pi-A$ scattering data~\cite{DUET:2016yrf}. 

\subsection{Near-detector analysis}
\label{sec:nd280}

The near detector complex is used to measure the properties of the neutrino beam before it oscillates. These measurements allow for a reduction of the systematic uncertainties that affect event rates at SK. 

An extended likelihood fit as a function of the reconstructed muon momentum and outgoing angle measured at ND280 is performed to constrain the (anti)neutrino flux and cross section modeling. Prior constraints are included as penalty terms. A total of 18 samples of \numu and \numub charged-current (CC) interactions with vertices in either of the FGDs are employed in this fit. Their selection is optimized to maximize the sensitivity of ND280 to different features of the (anti)neutrino spectra. Event selections are based on the requirement that the highest-momentum track is compatible with the muon hypothesis according to the TPC particle identification. This track is required to be negatively charged if the selection is performed in $\nu$-mode, but either positively or negatively charged in \nubar-mode to also identify the relatively large \numu background component of the \nubar-mode.
As in the previous analysis~\cite{PhysRevD.103.L011101}, in $\nu$-mode the sample of \numu CC interactions is further split into three subsamples according to the pion multiplicity in the final state: CC events without reconstructed pions (\CCzeropi), with one reconstructed positively-charged pion (\CConepiplus), and all remaining CC events (\CCother). In \nubar-mode, thanks to the increased statistics, we moved from a selection based on the track multiplicity to one that matches the selection adopted in $\nu$-mode. Such improvement was possible for both \numub and \numu background components, resulting in six \nubar-mode samples for each FGD. The main difference is related to the selection of \numub CC events with one reconstructed negatively-charged pion. They are identified by employing the particle identification capabilities of the TPC and FGD, and tagging the Michel electron produced in the $\pi \rightarrow \mu \rightarrow e$ decay chain. Since negatively-charged pions are more likely to be absorbed in the material of the FGD, if a Michel electron is tagged, the associated pion in 63\% of the cases is positively charged. The detector response is evaluated using dedicated control samples as detailed in Ref.~\cite{T2K:2015sqm}. Compared with previous analyses, pion secondary interactions (SI) are simulated using the semiclassical cascade model in NEUT, in place of the model used in previous analyses from GEANT4. The model was tuned to $\pi^{\pm}$-nucleus scattering data mentioned previously, which improved the agreement with data, reducing the systematic error associated with pion SI. 

Once the likelihood fit is performed, we calculate the $p$-value to quantify the ability of the best-fit point to describe the data, i.e.\ the probability of observing an outcome as or more extreme than data according to the model. It is computed as the fraction of fits for which the computed $\chi^2$ when varying the model is greater than the one computed for the fit to the data. We define $p$-values below 5\% as indicating a significant disagreement with the model. Over 895 variations of our model, we find a $p$-value of 74\%, much larger than this threshold. The result of the near-detector analysis is parametrized as a multivariate Gaussian constraint in the analysis employed to extract the oscillation parameters (\thatm, \dmsqtwothree) and (\thatmb, \dmsqtwothreeb).

\subsection{Far-detector event selection}
\label{sec:skevt}

This analysis uses two muonlike event samples at the far detector; one with the beam in $\nu$-mode and one with the beam in \nubar-mode. This allows for the oscillations of \numu and \numub to be measured separately despite the inability of SK to distinguish negatively charged and positively charged muons. The wrong-sign background in \nubar-mode is constrained by the $\nu$-mode samples by performing a combined analysis of the $\nu$- and \nubar-mode samples. 

Charge and timing information from the SK PMTs are used to reconstruct the vertex position, momentum, and particle identification (PID) of events inside the detector. Particles are identified by their Cherenkov ring profiles. Due to their larger mass, muons are more resilient to scattering, resulting in clear rings with well-defined edges. In contrast, electrons scatter more and produce electromagnetic showers, resulting in rings with diffuse edges. The reconstruction algorithm~\cite{Super-Kamiokande:2019gzr} also counts Michel electrons by identifying delayed hit timing clusters.

The samples used in this analysis, referred to as 1R$\mu$, select for reconstructed events with one muonlike ring and no other rings, and 0 or 1 delayed Michel electrons. The number of predicted (postnear-detector analysis) and observed events for both 1R$\mu$ samples are shown in \cref{tab:1rmu}. Note that the number of \nubar-mode 1R$\mu$ data events differs from the previous analysis described in Ref.~\cite{PhysRevD.103.L011101} due to updated data processing at SK, as described in Ref.~\cite{T2K:2023smv}. The increased exposure reduced the statistical uncertainty on the number of $\nu$-mode 1R$\mu$ events by 13\%, resulting in 5.6\%.

\begin{table}[h!]
	\caption{Number of predicted events and data events selected for both 1R$\mu$ samples. The predictions are calculated assuming $\dmsqonetwo=7.53\times10^{-5}\:\text{ eV}^2$, $\dmsqtwothree=2.509\times10^{-3}\:\text{ eV}^2$, $\ssqthtwothree=0.528$, $\ssqthonetwo=0.307$, $\ssqthonethree=0.0218$, $\deltacp=-1.601$, Earth matter density of $2.6 \:\text{ g} \text{ cm}^{-3}$, and normal mass ordering.}
	\label{tab:1rmu}
	\centering
	\begin{tabular}{c|c|c}
    	\toprule
    	Sample              & Prediction & Data \\
    	\midrule
    	$\nu$-mode 1R$\mu$  & 345.3      & 318 \\
    	\nubar-mode 1R$\mu$ & 135.2      & 137 \\
		\bottomrule
    \end{tabular}

\end{table}

\subsection{Impact of systematic uncertainties}

The systematic uncertainties we include in this analysis are associated with neutrino beam flux modeling, neutrino interaction cross section modeling, and detector response. The first two sources of systematic uncertainties are constrained by fitting our model to the ND280 near-detector data as described in \cref{sec:nd280}. The systematic uncertainties constrained by the near detector are included as input constraints in the far-detector analysis. \cref{tab:syst_errors} shows the contribution to the total relative uncertainty from each source of the systematic error on the predicted number of events in each SK sample. Both $\nu$- and \nubar-mode are reduced from 12\% to 3 and 4\% respectively thanks to the near-detector analysis. Some cross section systematics are not constrained by the near detector. The larger relative error on the 1R$\mu$ \nubar sample is mainly due to the large uncertainty in low-energy pion-production modeling. The far detector systematic error (SK det.) includes uncertainties in ring counting efficiencies, event selection, fiducial volume, secondary particle interactions, and photonuclear effects. 

Compared with the previous analysis, the total systematic error was reduced by 45\% and 9\%, for the 1R$\mu$ $\nu$-mode and $\nubar$-mode event rates respectively. As expected, this improvement is driven by the new flux tuning and new neutrino interaction modeling that are reduced overall by 36\% and 21\% for the two samples. The total systematic error must be compared with the statistical uncertainty which is 5.6\% and 8.5\%.

\begin{table}[htb!]
	\caption{Uncertainties on the number of events in each SK sample broken down by error source after the near-detector analysis. The first two rows show the uncertainties when flux and crosssection systematics (constrained by the near detector) are propagated without correlation, whereas the third (Flux+Xsec) has smaller uncertainties due to the anticorrelations in the near-detector analysis, and corresponds to what is used in the analysis. ``SK det." includes uncertainties from the SK detector response.}
	\label{tab:syst_errors}
	\centering
\begin{tabular}{l | c | c}
		\toprule
        Error source (units: \%) & 1R$\mu$ $\nu$-mode & 1R$\mu$ $\nubar$-mode \\ 
		\midrule\midrule
        Flux                   &  2.9 & 2.8 \\
        Xsec (ND constrained)  &  3.1 & 3.0 \\
		\midrule
        Flux+Xsec (ND constr.) &  2.1 & 2.3 \\
        SK-only Xsec           &  0.6 & 2.5 \\
        SK det.                &  2.1 & 1.9 \\
		\hline
        \textbf{Total}         &  3.0 & 4.0 \\
\hline\hline
	\end{tabular}
\end{table}

\subsection{Oscillation analysis}

The oscillation probabilities are calculated using a slight modification of the 3-flavor PMNS oscillation framework. The \numu survival probability, not including the matter effect for simplicity, is approximately given by
\begin{equation}\label{eqn:survprob}
\begin{aligned}
P\big(\,\parenbar{\nu}_{\mu}&\rightarrow \parenbar{\nu}_{\mu}\big) \simeq 1-(\cos^{4}\theta_{13}  \sin^{2}{2\parenbar{\theta}_{23}} \\
&\; + \sin^{2}2\theta_{13}  \sin^{2}{\parenbar{\theta}_{23}})
\times \mathrm{sin}^{2}\bigg(\frac{\Delta {\parenbar{m}}^{\;2}_{32} L}{4E}
\bigg)
\end{aligned}
\end{equation}
where the barred parameters correspond to muon antineutrino oscillations. The standard PMNS formalism is recovered when ($\sin^2\theta_{23}$, $\Delta m^{2}_{32}$) $=$ ($\sin^{2}\overline{\theta}_{23}$, $\Delta \overline{m}^{2}_{32}$). Note that the full \numu survival probability is employed in the analysis. In the oscillation analysis, neutrino and antineutrino parameters are varied independently and fitted simultaneously to data by minimizing the combined negative log-likelihood $-\ln \mathcal{L} = \sum_i \big(N_i^{\text{exp}}-N_i^{\text{obs}}+N_i^{\text{obs}}\times \ln(N_i^{\text{obs}}/N_i^{\text{exp}})\big)$ calculated for both muon neutrino and muon antineutrino samples binned in reconstructed neutrino energy and muon scattering angle, where $N_i^\text{exp}$ is the number of predicted events in the $i\text{th}$ bin and $N_i^\text{obs}$ is the number of observed events.
All systematic uncertainties and  other oscillation parameters, such as $\sin^22\theta_{13}$ and $\deltacp$, are treated as nuisance parameters and are marginalized over according to their assigned priors. This marginal likelihood is used to construct confidence intervals using the fixed $\Delta\chi^{2}$ method\cite{Wilks:1938dza}. Since the $\mu$-like samples are not sensitive to neutrino mass ordering or $\sin^22\theta_{13}$, we assume normal ordering in this analysis and constrain $\sin^22\theta_{13}$ by the Ref.~\cite{ParticleDataGroup:2018ovx} value from reactor experiments. As the survival probability from Eq.~\eqref{eqn:survprob} is symmetric in the sign of $\pm(\cos^2\theta_{13} \sin^2\parenbar{\theta}_{23} - 1/2)$, the constraints on $\sin^2\parenbar{\theta}_{23}$ will be symmetric about $0.5/\cos^2\theta_{13} \approx 0.511$; in the standard PMNS formalism analysis this symmetry is broken by the inclusion of \nue and $\bar{\nu}_e$~samples. A flat prior is used for \deltacp. The robustness of the analysis is assessed by repeated tests using a variety of simulated data sets with alternative interaction models. The bias on the parameters of interest is estimated as well.

\section{Results and Discussion} 
\label{sec:results}

\begin{figure}[ht!] 
    \centering
    \includegraphics[width=0.45\textwidth]{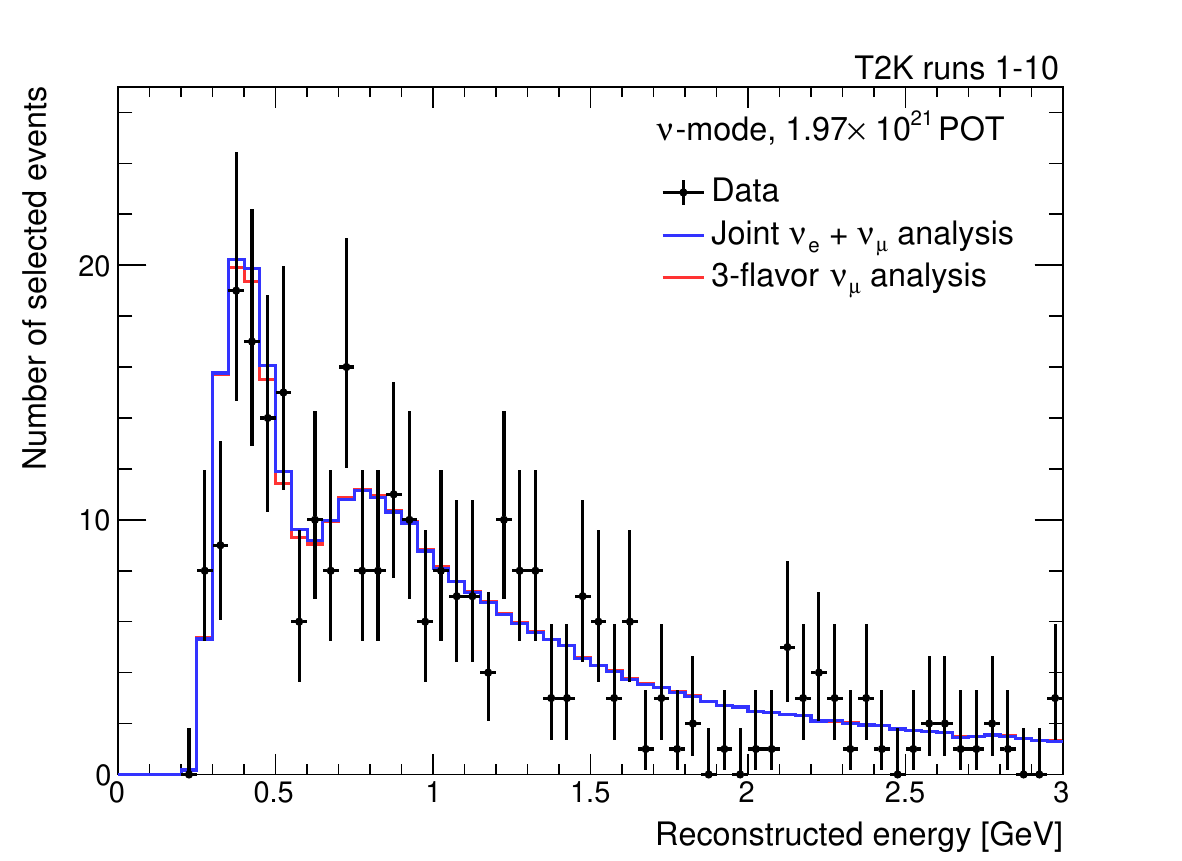}
    \includegraphics[width=0.45\textwidth]{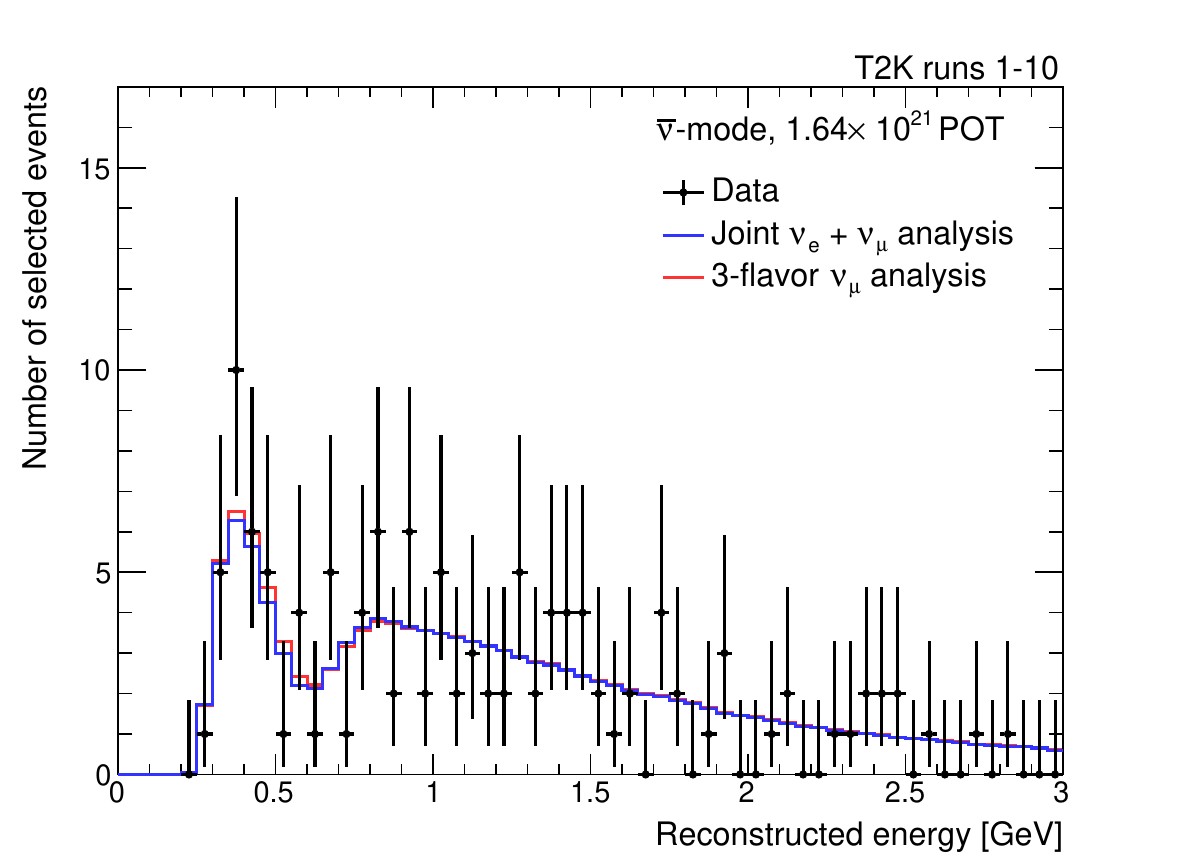}
    \caption{The reconstructed neutrino energy distributions for neutrino (top) and antineutrino (bottom) mode 1R$\mu$ samples. The lines show the predicted number of events under two hypotheses: ``Joint $\nu_{e}$ + $\nu_{\mu}$ analysis'' uses the best-fit values from a joint analysis of the PMNS model to electronlike and muonlike samples\cite{T2K:2023smv}, ``3-flavor $\nu_{\mu}$ analysis'' (this analysis) uses the best-fit from the analysis reported here. The error bars indicate the statistical uncertainties. }
    \label{fig:data_pred_bestfit}
\end{figure}

The reconstructed energy distributions for $\nu$-mode and \nubar-mode 1R$\mu$ samples for data taken from January 2010 to February 2020 (run 1--10) and the best-fit predictions are shown in \cref{fig:data_pred_bestfit}. The results of the three-flavor analysis using both electronlike and muonlike samples as described in Ref.~\cite{T2K:2023smv} are also shown for comparison. In both cases, the prediction and data agree within the statistical uncertainties indicated by the error bars.

The best-fit values obtained for oscillation parameters describing neutrino oscillations are $\sin^2\theta_{23} = 0.47^{+0.11}_{-0.02}$ and $\Delta m^{2}_{32} = 2.48^{+0.05}_{-0.06}\times 10^{-3}\,\:\si{\square\electronvolt}$ and those describing antineutrino oscillations are $\sin^2\overline\theta_{23} = 0.45^{+0.16}_{-0.04}$ and $\Delta \overline m^{2}_{32} = 2.53^{+0.10}_{-0.11}\times 10^{-3}\,\:\si{\square\electronvolt}$. The best-fit values for both neutrino and antineutrino oscillations agree within the uncertainties. 

Based on the robustness checks, the bias on $\Delta m^{2}_{32}$ introduced by the limited flexibility of the neutrino interactions model for $\nu$-mode (\nubar-mode) is estimated to be 1.40 (1.55)$\times 10^{-5}\:\si{\square\electronvolt}$, which is accounted for in the analysis by smearing the $\Delta\chi^{2}$ contour with additional Gaussian uncertainty. As for the analysis in Ref.~\cite{T2K:2023smv}, the biggest bias was observed using an alternative model for pion secondary interactions. No bias is observed on the other oscillation parameters. 

\begin{figure}[ht!] 
    \centering
    \includegraphics[width=0.45\textwidth]{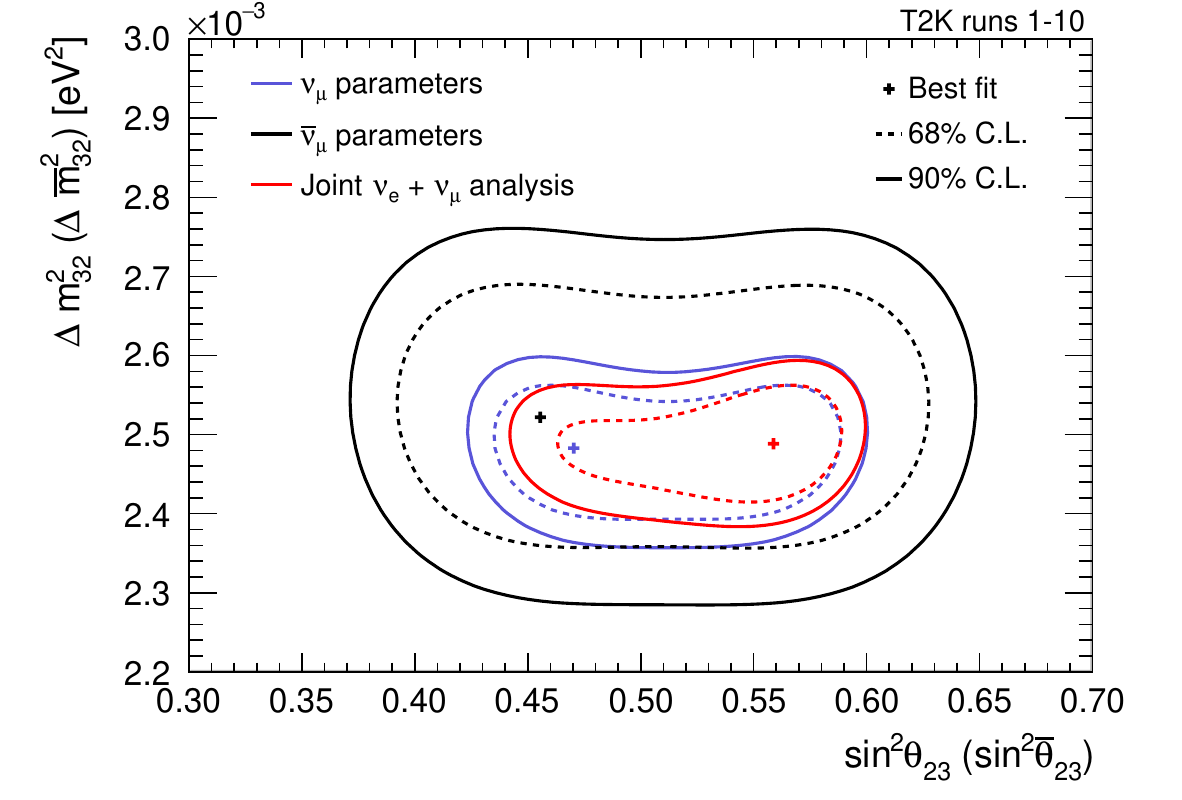}
    \caption{Confidence regions of ($\sin^2\theta_{23}, \Delta m^{2}_{32}$) for neutrinos and their barred parameters for antineutrinos. Corresponding regions from the standard PMNS formalism analysis~\cite{T2K:2023smv} including $\nue$ samples are also shown for comparison.}
    \label{fig:run1_10}
\end{figure}

In \cref{fig:run1_10} we compare the constraints on $\sin^{2} \overset{\scriptscriptstyle(-)}{\theta}_{23}$ and $\Delta \overset{\scriptscriptstyle(-)}{m}{}^{2}_{32}$ coming from the three-flavor analysis to muonlike samples and the joint analysis to both electronlike and muonlike samples~\cite{T2K:2023smv}. Since the parameters for $\numu$ and $\bar{\nu}_{\mu}$ are compatible, this analysis does not provide indication of new physics. The $\numu$-only analysis results are not sensitive to the $\theta_{23}$ octant due to the lack of electronlike samples in the analysis.

\begin{figure}[ht!]
    \centering
    \includegraphics[width=0.45\textwidth]{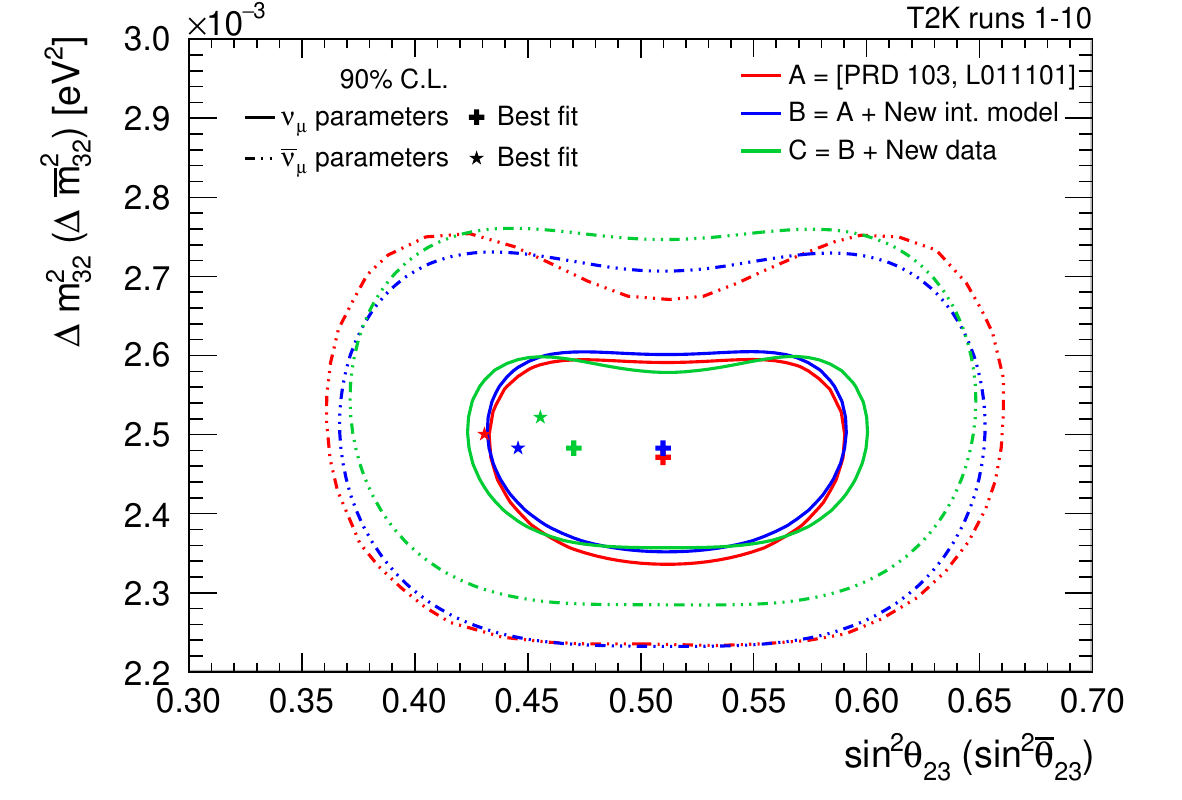}
    \caption{Comparison of the 90\% confidence level in ($\sin^2\theta_{23}, \Delta m^{2}_{32}$) and their barred counterparts for antineutrinos (dot-dashed lines) to those obtained in the previous analysis (red line), an intermediate step showing the contribution of the updated analysis model (blue line), and the final results of this analysis including new SK neutrino mode data and updated data processing (green line).}
    \label{fig:evolution_datafit}
\end{figure}

\cref{fig:evolution_datafit} shows a comparison to the results obtained in the previous analysis and an intermediate step to show the contribution of the updated analysis model. The ana\-ly\-sis model is found to change the shape of the antineutrino parameter contours, whereas the new data at SK improve the background constraint and move the neutrino parameters away from maximal mixing. The new SK data also move the antineutrino parameters to slightly larger values, but compatible, of $\Delta \bar{m}^{2}_{32}$ and $\sin^{2}\bar{\theta}_{23}$, which is also affected by the updated data processing.

\section{Conclusions}
\label{sec:conclusions}

We have presented the results from the muon (anti)neutrino oscillation analysis to T2K data corresponding to a total of 3.6 $\times$ 10$^{21}$ POT taken in neutrino and antineutrino mode. The predictions for each SK sample are based on the constraints provided by the near-detector analysis. We conclude that the measurements of the parameters describing the oscillations of muon neutrinos and antineutrinos are compatible with the three-flavor prediction and provide no indication of new physics. The data related to this work can be found in Ref.~\cite{datarelease}.

\section*{Acknowledgments}

We thank the J-PARC staff for superb accelerator performance. We thank the CERN NA61/SHINE Collaboration for providing valuable particle production data. We acknowledge the support of MEXT, JSPS KAKENHI (JP16H06288, JP18K03682, JP18H03701, JP18H05537, JP19J01119, JP19J22440, JP19J22258, JP20H00162, JP20H00149, JP20J20304) and bilateral programs(JPJSBP120204806, JPJSBP120209601), Japan; NSERC, the NRC, and CFI, Canada; the CEA and CNRS/IN2P3, France; the DFG (RO 3625/2), Germany; the INFN, Italy; the Ministry of Education and Science(2023/WK/04) and the National Science Centre (UMO-2018/30/E/ST2/00441 and UMO-2022/46/E/ST2/00336 ), Poland; the RSF19-12-00325, RSF22-12-00358, Russia; MICINN (SEV-2016-0588, PID2019-107564GB-I00, PGC2018-099388-BI00, PID2020-114687GB-I00) Government of Andalucia (FQM160, SOMM17/6105/UGR) and the University of Tokyo ICRR's Inter-University Research Program FY2023 Ref. J1, and ERDF funds and CERCA program, Spain; the SNSF and SERI (200021\_185012, 200020\_188533, 20FL21\_186178I), Switzerland; the STFC and UKRI, UK; and the DOE, USA. We also thank CERN for the UA1/NOMAD magnet, DESY for the HERA-B magnet mover system, the BC DRI Group, Prairie DRI Group, ACENET, SciNet, and CalculQuebec consortia in the Digital Research Alliance of Canada, GridPP and the Emerald High Performance Computing facility in the United Kingdom, and the CNRS/IN2P3 Computing Center in France. In addition, the participation of individual researchers and institutions has been further supported by funds from the ERC (FP7), ``la Caixa'' Foundation (ID 100010434, Fellowship Code No. LCF/BQ/IN17/11620050), the European Union's Horizon 2020 Research and Innovation Programme under the Marie Sklodowska-Curie Grant Agreement No. 713673 and No. 754496, and No. H2020 Grant No. RISE-GA822070-JENNIFER2 2020 and No. RISE-GA872549-SK2HK; the JSPS, Japan; the Royal Society, UK; French ANR Grant No. ANR-19-CE31-0001; the SNF Eccellenza Grant No. PCEFP2\_203261; and the DOE Early Career programme, USA. For the purposes of open access, the authors have applied a Creative Commons Attribution licence to any Author Accepted Manuscript version arising.

\bibliography{apssamp}

\end{document}